\documentclass[
	11pt,				
	oneside,			
	a4paper			
	]{article}

\usepackage{lmodern}			
\usepackage[T1]{fontenc}		
\usepackage[utf8]{inputenc}	
\usepackage{indentfirst}		
\usepackage{nomencl} 			
\usepackage{color}				
\usepackage{graphicx}			
\usepackage{microtype} 			
\usepackage{amsmath, amsfonts}
\usepackage{physics}
\usepackage{placeins}
\usepackage[style=nature, sorting=none]{biblatex}
\usepackage{comment}
\usepackage{lineno}
\usepackage{soul}

\title{Bidimensional Symplectic Maps}

\author{Felipe G. Souza$^1$, Gabriel C. Grime$^1$, Iberê L. Caldas$^1$ \\
\small 1. University of São Paulo, Institute of Physics, \small São Paulo-SP, Brazil, \\
}
\date{\today}

\definecolor{blue}{RGB}{41,5,195}

\makeatletter

\usepackage[colorlinks]{hyperref}
\hypersetup{
    colorlinks=true,
    linkcolor=blue,
    filecolor=blue,      
    urlcolor=blue,
    citecolor=blue,
}

\usepackage{caption}
\usepackage{subcaption}

\usepackage[margin=3cm]{geometry}

\begin{document}

\frenchspacing 


%
%
\maketitle

\begin{abstract}
\noindent Symplectic maps can provide a straightforward and accurate way to visualize and quantify the dynamics of conservative systems with two degrees of freedom. These maps can be easily iterated from the simplest computers to obtain trajectories with great accuracy. Their usage arises in many fields, including celeste mechanics, plasma physics, chemistry, and so on. In this paper we introduce two examples of symplectic maps, the standard and the standard non-twist map, exploring the phase space transformation as their control parameters are varied. 
\\ \textbf{Keywords:} Symplectic Maps, Standard Map, Standard Non-Twist Map, Conservative Systems.
\end{abstract}

\section{Introduction}
 
The phase space of a dynamical system represents all its possible states. Starting from an initial condition, trajectories along the phase space points represent an evolution of the system in time \cite{lichtenberg,ozorio}. 

In particular, for two degrees of freedom systems, the time progression of the flow of trajectories in phase space can be studied within a bidimensional surface as the phase paths intersect it successively \cite{lichtenberg}. These surfaces are known as Poincaré sections and can be generalized for a $2N$ degree of freedom system. The symplectic maps, explained in detail in the following section, can be approximately depicted as Poincaré sections of two degrees of freedom quasi-integrable systems. 
 
This article is for students of physics, with the goal of inviting them to delve deeper into the subject of area-preserving maps by introducing fundamental concepts of symplectic maps using two examples: the standard and standard non-twist maps. They are paradigmatic systems, serving the purpose of introducing symplectic maps.

The standard map is a family of area-preserving maps with a single parameter $k$ that expresses the disturbance of the system. It was introduced by Chirikov \cite{chirikov} establishing its universality and numerous applications, as the description of the dynamics of magnetic field lines, and independently by Taylor \cite{taylor}, modelling magnetic fields in plasmas. Another significant result for the standard map was obtained by Greene \cite{greene1979}, involving a method to determine the transition to chaos.

First introduced by del Castillo Negrete and Morrison in 1993 \cite{del-Castillo-Negrete1993} while studying the chaotic transport in shear flow and described with further detail by Wurm, Apte, Fuchss, and Morrison in 2005 \cite{wurm2005}, the standard non-twist map is defined as a map that locally violates the twist condition. 

This violation of the twist condition leads to some unique properties of physics interest, concerning transport barriers and the map's topology, such as the twin islands chain appearance and the reconnection process between others. The curve where the violation happens, the shearless curve, is destroyed for specific sets of parameters. This set can be obtained with the standard non-twist map breakup diagrams, where each point represents a pair of parameters.

The standard and standard non-twist maps can be applied to model a large variety of physical theories, e.g., magnetic fields in tokamaks \cite{portela2008, morrison2000, oda1995, caldas2012} and stellarators \cite{hayashi1995} (plasma physics), planetary orbits \cite{kyner1968,moser2001} and stellar pulsations \cite{munteanu2002} (astronomy), sheared flows \cite{del-Castillo-Negrete1993, del-Castillo-Negrete2000}.

This paper is organized as follows. In section 2, we start with a review of the main concepts of quasi-integrable conservative systems in order to introduce the symplectic maps. Sections 3 and 4 are dedicated to the example maps, exploring the general properties of the symplectic maps as well as some unique from the standard and standard non-twist maps. Section 5 presents the parameter space as a tool to investigate the shearless curve breakup.

\section{Symplectic Maps}

Essentially, a dynamical system can be described by its phase space using the Hamiltonian framework. For example, a $4$-dimensional phase space will be formed of the generalized positions $\{q_1,q_2\}$ and momentum $\{p_1,p_2\}$ showing all possible states of the system. These two degrees of freedom \emph{Hamiltonian dynamical systems} are fundamentally symplectic\footnote{The term symplectic comes from the greek word \emph{sumplektikos} and it means intertwined.} \cite{abud2013}, this is, they preserve both volume and infinitesimal areas in phase space, according to the \emph{Liouville's Theorem} \cite{meiss1992}.

Dynamical systems with two degrees of freedom may become hard to deal with, given their four-dimensional phase space. In order to simplify the analysis of the phase spaces, Poincaré introduced the concept of \emph{Poincaré sections}, as they are known nowadays. These sections are cross-sectional surfaces to the flux of trajectories in phase space \cite{poincare1893}. Thereby, for a time-independent Hamiltonian system of two degrees of freedom, the Poincaré sections allow us to simplify the problem, returning us to a two-dimensional system. 

\begin{figure}[htb]
    \centering
    \includegraphics[width=6.82cm]{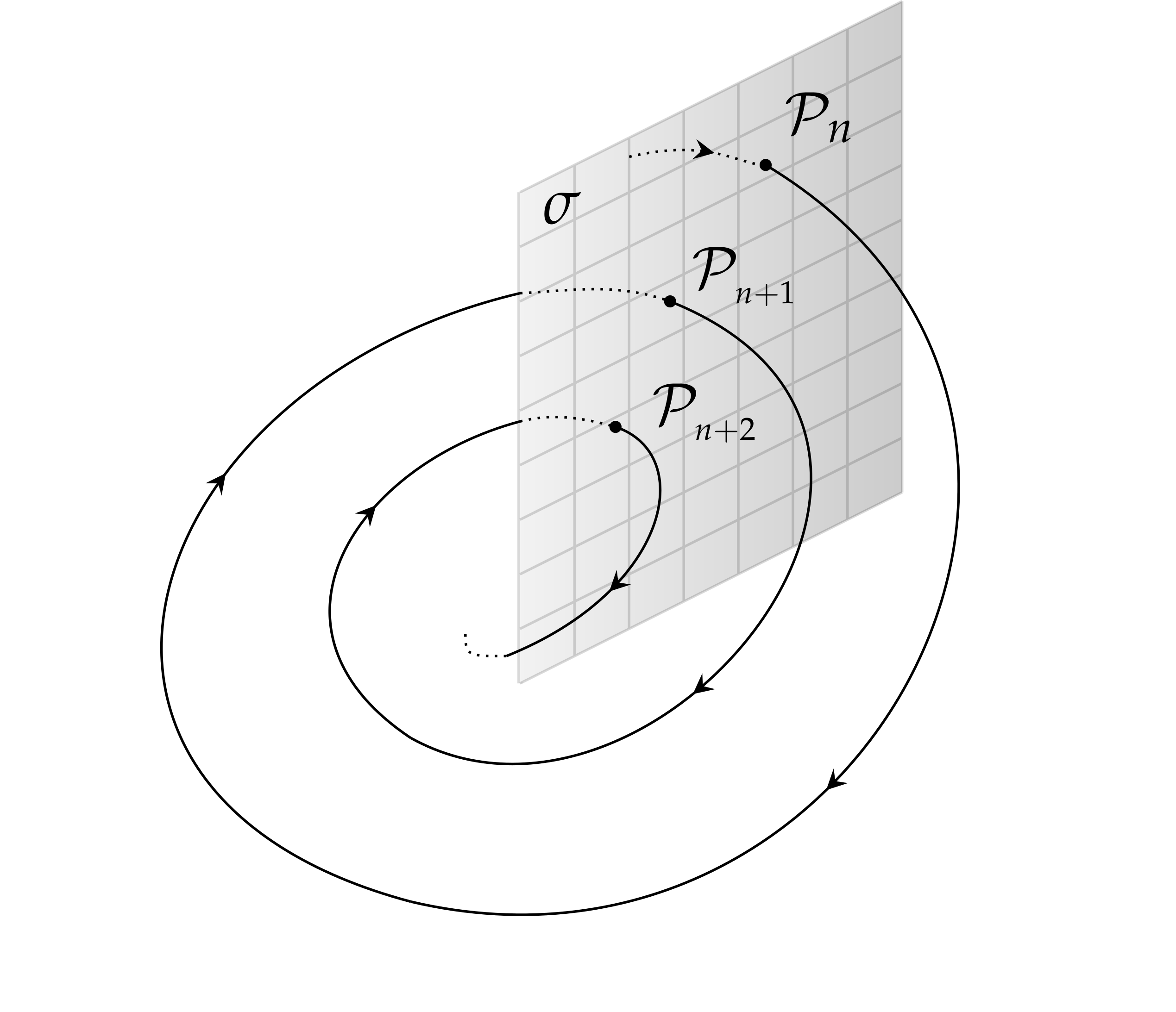}
    \caption{Phase diagrams passing through a 2-dimension Poincaré Section $\sigma$ and its iterated intersection points $\mathcal{P}$.}
    \label{draw}
\end{figure}

For the nth intersection of the trajectory of a dynamical system with a Poincaré section in phase space being $\mathcal{P}_n$, the next point of intersection given the time evolution will be $\mathcal{P}_{n+1}$. We define a \emph{Poincaré map} as the discrete dynamics given by a matrix $\mathcal{M}$ in such a way that $\mathcal{P}_{n+1} = \mathcal{M}\mathcal{P}_n$ \cite{lichtenberg}, being each point $\mathcal{P}$ placed only in a single direction. Given this framework, we define a discrete \emph{orbit} as a sequence of points $\{\mathcal{P}_0,\dots,\mathcal{P}_n,\mathcal{P}_{n+1},\dots\}$ such that $\mathcal{P}_{n+1} = \mathcal{M}\mathcal{P}_n$ \cite{meiss1992}.

An indispensable concept for the understanding of the symplectic maps is the \emph{integrability}. A two-degrees-of-freedom system is said integrable if there exists exactly two independent \emph{integrals of motion} \cite{lichtenberg,reichl} (essentially the degrees of freedom of the system must match the number of integrals of motion). These so-called integrals of motion are functions $\mathcal{F}(q_n,p_n), \ n=1,2$ that are constants along a trajectory in phase space \cite{nivaldo}. 

Let's now consider the following perturbed Hamiltonian \cite{lichtenberg}
\begin{equation}
\label{Perturbed Hamiltonian}
    \mathcal{H}(\theta,\mathcal{J}) = \mathcal{H}_0(\mathcal{J}) + \epsilon\mathcal{H}_1(\theta,\mathcal{J}),
\end{equation}
where ($\theta$,$\mathcal{J}$) represents the action-angle variables of the integrable Hamiltonian $\mathcal{H}_0$\footnote{The nth action variable is defined as $\mathcal{J}_n = \frac{1}{2\pi}\oint p dq, \ n = 1,\dots,N$ and the nth angular variable as $\theta_n = \pdv{W}{\mathcal{J}_{n}}, \ n = 1,\dots,N$ with the line integral along one period and $W$ being the complete integral of the Hamiltonian system \cite{nivaldo}}. The term $\mathcal{H}_1$ expresses the non-integrable part of the Hamiltonian and $\epsilon$ is the amplitude of perturbation. 

From Kolmogorov-Arnold-Moser's (KAM) theorem, for frequencies $\Omega = \pdv{\mathcal{H}_0}{\mathcal{J}}$ sufficiently irrational and for $\epsilon\ll 1$, most of the invariant tori will become slightly deformed but preserving their former properties \cite{lichtenberg,reichl,ozorio}. In other words, for small perturbations, the system has just a slight change in most of its phase space. However, this kind of system has also chaotic orbits around the so-called hyperbolic points, which will be described later.

For these quasi-integrable systems, the regular orbits of irrational frequencies are presented in Poincaré sections by curves or \emph{invariant tori} \cite{lichtenberg} (topologically they match with the definition of a torus). These curves are usually referred to as orbits or simply tori. However, in the limit $\epsilon\gg1$ (overwhelming KAM's theorem validity), more tori become broken implying global chaos. Their presentation in Poincaré sections is just of densely distributed points.

In the same way, the KAM theorem states the quasi-integrability for irrational frequencies, but it fails to provide the same conclusion for the rational ones. Complementing KAM's theorem, \emph{Poincaré-Birkhoff theorem} states that, considering a torus with rational frequency $\Omega = r/s, \ r,s\in \mathbb{Z}$, from the perturbation $H_1$ creates an even number $2ms,\ m\in \mathbb{N}$, of periodic points. From these points, half of them are stable, expressing the labeled \emph{elliptical points} and the other half unstable, called \emph{hyperbolic points} \cite{birkhoff1913,aguiar}. Both of them will be described in the following section.

A \emph{periodic orbit} of period $s$ is defined as $\mathcal{M}^s(\theta_n,\mathcal{J}_n) = (\theta_{n} + r,\mathcal{J}_{n}), \ \forall\ n$, as a consequence of this statement, their frequencies are rational numbers $\Omega = r/s, \ r,s\in \mathbb{Z}$ \cite{lichtenberg}. Different from the quasi-periodic orbits, from Poincaré-Birkhoff's theorem the rational frequency implies periodic distributions of the iterated points, presenting themselves in Poincaré sections in closed ellipses.

The time evolution in \emph{quasi-integrable conservative dynamical systems} can be discrete (map) or continuous (differential equations). The first one gives rise to the \emph{symplectic maps}, defined as area-preservative conservative systems, described by Poincaré sections of time-independent bidimensional Hamiltonians \cite{lichtenberg}, given by recurrence relations of the approximated form \cite{lichtenberg}:
\begin{equation}
\label{General representation of the iteration form of symplectic maps}
\begin{cases}
    \mathcal{J}_{n+1} = \mathcal{J}_{n} + \epsilon f(\theta_n,\mathcal{J}_{n+1}) \\
    \theta_{n+1} = \theta_{n} + \Omega(\mathcal{J}_{n+1}) + \epsilon g(\theta_n,\mathcal{J}_{n+1})
\end{cases}
\end{equation}
for $n \in \mathbb{N}$ being the nth iteration, $f$ and $g$ analytic periodic functions in $\theta$ and $\Omega$ the unperturbed frequency.

\begin{figure*}[htb]
    \centering
    \begin{subfigure}[b]{.48\textwidth}
        \centering
        \includegraphics[width=\textwidth]{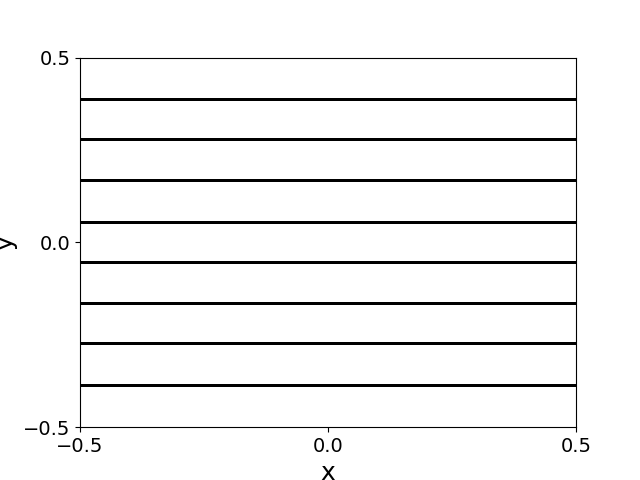}
        \caption{$k = 1\times 10^{-4}$.} 
        \label{k = 10(-4)}
    \end{subfigure}
    \hfill
    \begin{subfigure}[b]{.48\textwidth}
        \centering 
        \includegraphics[width=\textwidth]{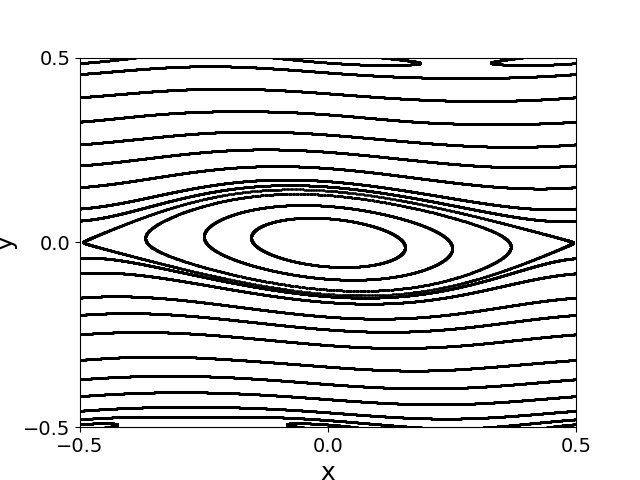}
        \caption{$k = 0.0318$.}  
        \label{k = 0.0318}
    \end{subfigure}
    \vskip\baselineskip
    \begin{subfigure}[b]{.48\textwidth}
        \centering 
        \includegraphics[width=\textwidth]{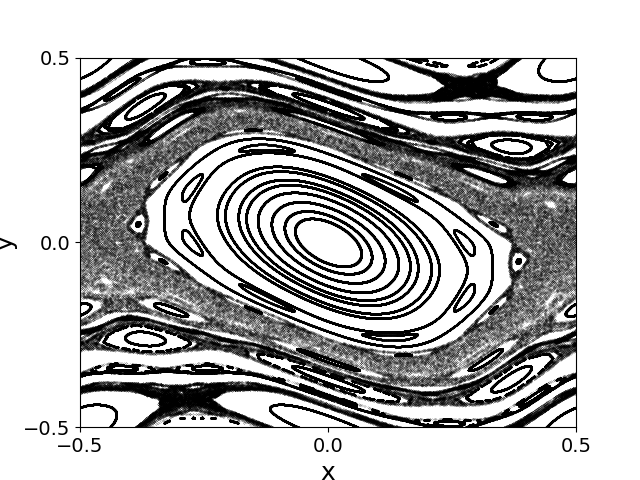}
        \caption{$k = 0.151$.}  
        \label{k = 0.151}
    \end{subfigure}
    \hfill
    \begin{subfigure}[b]{.48\textwidth}  
        \centering 
        \includegraphics[width=\textwidth]{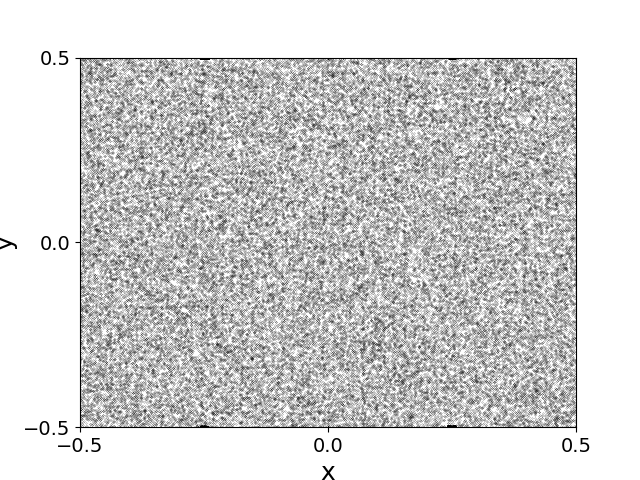}
        \caption{$k = 4$.}   
        \label{k = 4}
    \end{subfigure}
    \caption{Example of a sequence of standard maps increasing the value of the disturbance parameter $k\geq 0$.}
    \label{Standard Map: maps}
\end{figure*}

Since these maps are symplectic, the \emph{area-preserving} property must hold. For a standard symplectic map $(\mathcal{J}_{n+1},\theta_{n+1}) = \mathcal{M}(\mathcal{J}_{n},\theta_n)$ (matrix form) this characteristic is expressed as the determinant of its Jacobian matrix being unitary \cite{meiss1992}.

\begin{equation}
\mathrm{det} \left[\pdv{(\mathcal{J}_{n+1},\theta_{n+1})}{(\mathcal{J}_{n},\theta_n)}\right] = 1
\end{equation}

This is an important statement about these maps. Since it is a unique property of them, the symplectic maps can also be called \emph{area-preserving} \emph{maps}.

In the following sections, we will work with two examples of symplectic maps, the standard and the standard non-twist maps.

\section{The Standard Map}

Also known as  \emph{Chirikov-Taylor's map} the standard map is described by the recurrence equations \cite{chirikov,taylor},
\begin{equation}
\label{Standard Map Equations}
\begin{cases}
    y_{n+1} = y_n + k\sin(2\pi x_n)\\
    x_{n+1} = x_n - y_{n+1},\ \ \ \ \ \ \ \ \ \ \ \ \  \mbox{mod } 1
\end{cases}
\end{equation}
in which $x_n$ and $y_n$ represents the nth iteration of the variables $\theta_n$ and $\mathcal{J}_n$ respectively and $k\geq 0$ is a perturbation parameter. The value of $k$ is directly connected with the constant $\epsilon$ in Equation (\ref{Perturbed Hamiltonian}) and therefore as we increase it, the non-linearity of the

\begin{figure}[htb]
 \centering
 \begin{subfigure}[b]{.48\textwidth}
     \centering
     \includegraphics[width=\textwidth]{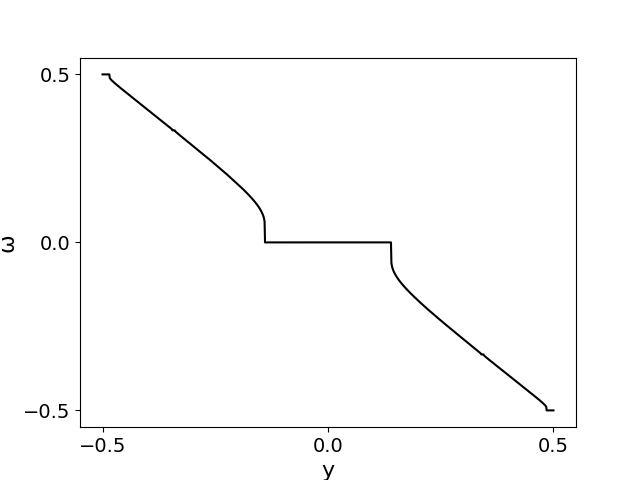}
     \caption{Winding number profile for $k = 0.0318$.}
     \label{RN k = 0.0318}
 \end{subfigure}
 \hfill
 \begin{subfigure}[b]{.48\textwidth}
     \centering
     \includegraphics[width=\textwidth]{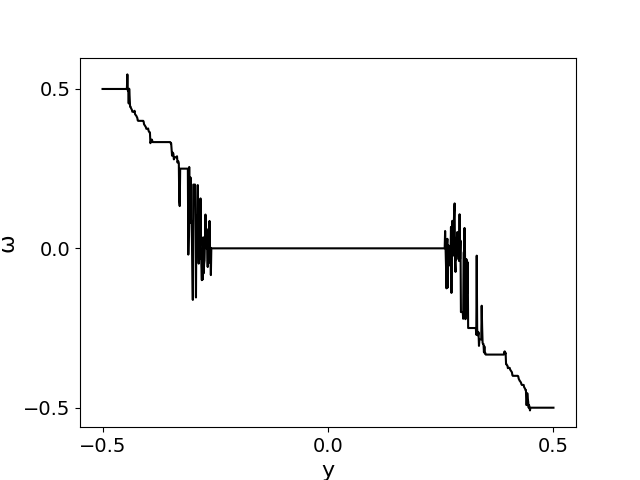}
     \caption{Winding number profile for $k = 0.151$.}
     \label{RN k = 0.151}
 \end{subfigure}
 \caption{Examples of winding number profiles increasing the disturbance parameter $k$.}
\label{Standard Map: winding number profiles}
\end{figure}

\noindent map raises, revealing eventually chaotic structures \cite{chirikov}.

The system is a twist map since it satisfies the \emph{twist condition} \cite{lichtenberg,reichl}:
\begin{equation}
    \label{Twist-Condition}
    \pdv{x_{n+1}}{y_n} \neq 0
\end{equation}
everywhere. This circumstance is the analog of the \emph{nondegeneracy condition} from Hamiltonian systems \cite{del-Castillo-Negrete1996}:
\begin{equation}
    \label{Non-Degeneracy Condition}
    \pdv[2]{H_0}{\mathcal{J}} \neq 0
\end{equation}
and it's a crucial statement for KAM's theorem applicability to the map.

From Fig. (\ref{k = 10(-4)}) we can see the standard map for the small value of the parameter $k = 1\times 10^{-4}$. This choice makes the non-linear term $\sin(2\pi x_n)$ a diminutive contribution to the Equations (\ref{Standard Map Equations}). This implies a near-linear behavior on the map thus making the phase space with predominant regular orbits. 

The almost horizontal lines in Fig. (\ref{k = 10(-4)}) are called \emph{invariant curves} or \emph{KAM curves} and each one of them represents quasi-periodic orbits in phase space. Each invariant curve is related to an initial condition $(x_0,y_0)$, in Fig. (\ref{k = 10(-4)}) only a few invariants are shown in order to keep the image clean.

If we increase the magnitude of the parameter to $k=0.0318$, as we can see in Fig. (\ref{k = 0.0318}), the non-linearity starts to take place. The invariant curves start to deform as stated by the KAM's theorem. It is also very prominent in the same figure the arising of \emph{islands} on the map that represent the periodic orbits in phase space, as predicted by the Poincaré-Birkhoff theorem. The central point of an island is a periodic elliptical point, surrounded by stable quasi-periodic motion around it \cite{reichl,greene1968}. The points on both sides of the islands in Fig.(\ref{k = 0.0318}) are the periodic  hyperbolic points, being unstable, with a small chaotic region around them \cite{abud2013,greene1968}.

Another important concept to extract from Fig. (\ref{k = 0.0318}) is the \emph{separatrix} curve. The separatrices are the invariant lines that almost connect the hyperbolic points. These curves, for quasi-integrable maps, are along the hyperbolic points, marking a boundary between two different behaviors from the phase space paths. In fact, in Fig. (\ref{k = 0.0318}), the separatrix traces the boundary between the invariant curves inside the islands and others outside.

For $k = 0.151$ in Fig. (\ref{k = 0.151}), a large number of islands appear, revealing \emph{chaotic regions} and thus the non-integrability of the system. For a high value of the parameter, as $k = 4$ for example (Fig. (\ref{k = 4})), the map gets completely chaotic, showing no sign of KAM curves and islands. The chaotic regions

\begin{figure}[htb]
 \centering
 \begin{subfigure}[b]{.48\textwidth}
     \centering
     \includegraphics[width=\textwidth]{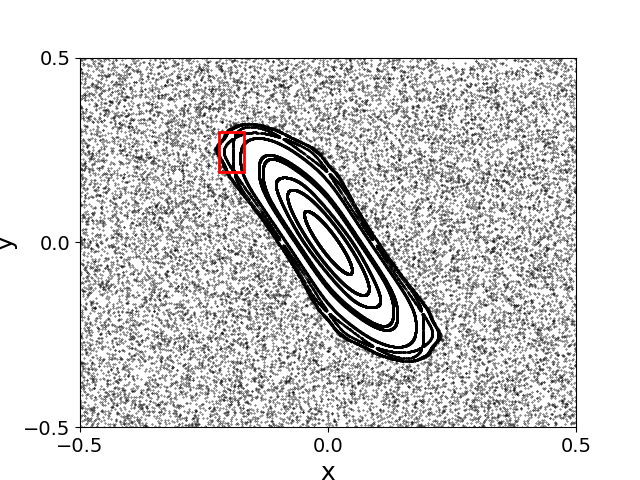}
     \caption{$k = 0.509$.}
     \label{k = 0.509}
 \end{subfigure}
 \hfill
 \begin{subfigure}[b]{.48\textwidth}
     \centering
     \includegraphics[width=\textwidth]{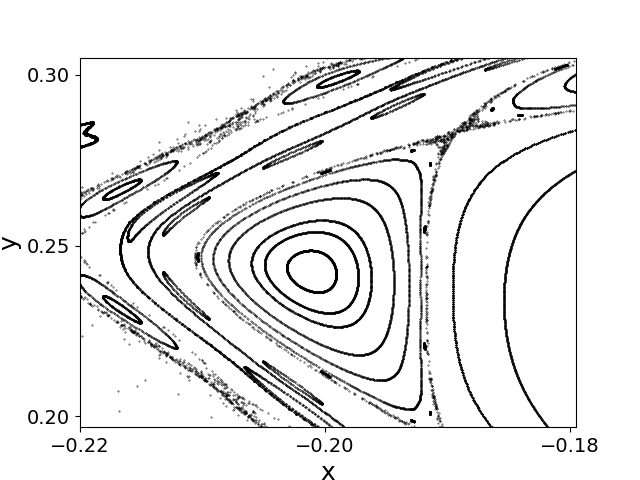}
     \caption{$k = 0.509$ map with zoom.}
     \label{k = 0.509 (zoom)}
 \end{subfigure}
 \caption{Example of the standard map fractal behavior. The red rectangle indicated in (\ref{k = 0.509}) is shown amplified in Fig. (\ref{k = 0.509 (zoom)}).}
\label{Standard Map: fractal behavior}
\end{figure}

\noindent shown in Figs. (\ref{k = 0.151}) and (\ref{k = 4}) originates from the hyperbolic points, having an unpredictable behavior in the long run due to the high sensitivity to the initial conditions. During the transition from Fig. (\ref{k = 0.151}) to (\ref{k = 4}), there is a critical value of $k$ from which the chaotic regions are no longer separated. The numeric calculation of this threshold value was made for the first time by John Greene \cite{greene1979}.

Chaotic maps can also have their topology and monotonicity obtained from \emph{winding number profiles}. These profiles are based on the definition of \emph{winding number} \cite{meiss1992,reichl} (also called \emph{rotation number})
\begin{equation}
\label{winding Number Definition}
    \omega = \lim_{n\to\infty}\frac{x_{n+1} - x_0}{n},
\end{equation}
provided the limit exists. Invariant curves and islands have a dense behavior as we can see in the following figures, thus having an existing winding number corresponding to each one of them. 

For $\epsilon = 0$, invariant curves (quasi-periodic orbits) have irrational winding numbers, while islands (periodic orbits) have their rotation numbers given by a rational number $\omega = r/s, \ r,s \in \mathbb{Z}$ (where $r$ is the number of revolutions in $\theta$ and $s$ is the period) \cite{lichtenberg}. 

The winding number profiles are $\omega\times y$ graphics that show how the winding numbers within an arbitrary curve act with respect to the coordinates of the vertical $y$-axis (For the standard map we will always be using the line $x = 0.5$ as the reference curve.). 

Looking at the winding number profile in Fig. (\ref{RN k = 0.0318}), by comparing the major island interval in the $y$-axis with the correspondent one in the map (\ref{k = 0.0318}), it becomes clear that the larger island in the middle has a constant winding number, forming the central plateau. Thus, having the winding number profile of a map enable us to identify islands along the chosen curve without looking at the map itself and obtain its rotation number. 

 As we can see in Fig. (\ref{RN k = 0.151}), a large amount of plateaus in the figure actually represents the numerous islands from the map (Fig. (\ref{k = 0.151})). However, a new element presents in the graphic, the irregular (clearly not continuous) dark oscillating lines in the winding number profile which can be identified as places where the winding number does not exist (as the limits from Equation (\ref{winding Number Definition})). If we compare the $y$-axis intervals of these regions within the reference curve $x=0$ with the ones from its correspondent map, these regions evidently are chaotic. Thus from the winding number profile, irregular regions (places where $\omega$ does not exist) as the ones from Fig. (\ref{RN k = 0.151}) represent

\begin{figure*}[htb]
 \centering
 \begin{subfigure}[b]{.48\textwidth}
     \centering
     \includegraphics[width=\textwidth]{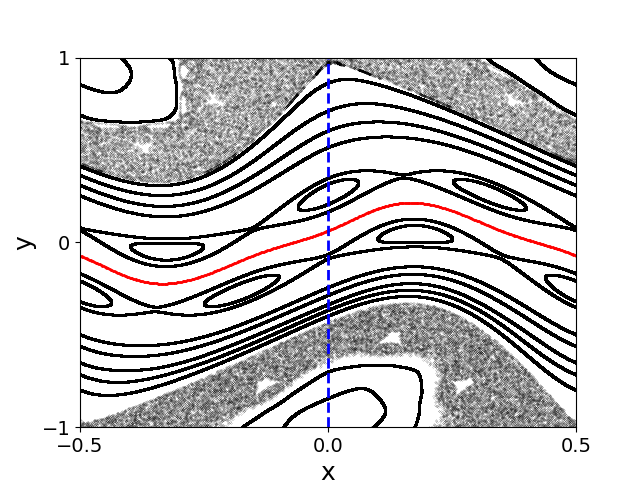}
     \caption{$a = 0.35$ and $b = 0.35$.}
     \label{a = 0.35}
 \end{subfigure}
 \hfill
 \begin{subfigure}[b]{.48\textwidth}
     \centering
     \includegraphics[width=\textwidth]{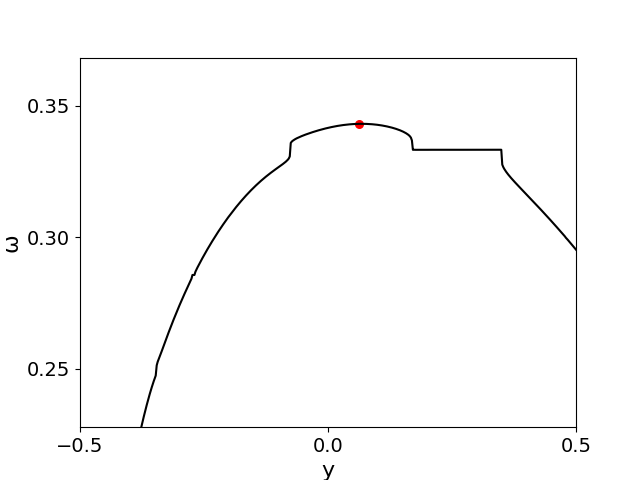}
     \caption{Winding number profile for a = 0.35 and b = 0.35.}
     \label{RN a = 0.35}
 \end{subfigure}
 \caption{The standard non-twist map and its nonmonotonic winding number profile as an example.}
\label{Non-Twist Map: general example}
\end{figure*}

\noindent chaotic regions on the map.

In Fig. (\ref{Standard Map: winding number profiles}) the \emph{monotonicity} of the standard map profiles is evident. A curve is said monotonic if is either strictly increasing or decreasing in an interval, thus matching the behavior seen in Fig. (\ref{Standard Map: winding number profiles}).

The standard map has an interesting property, it has a fractal behavior \cite{lichtenberg}. Generally, fractals are geometric patterns that keep their topological structures for arbitrarily small scales \cite{mandelbrot} In Fig. (\ref{k = 0.509}) (standard map for $k = 0.509$), if we choose to take a closer look inside the red rectangle, we'll get Fig. (\ref{k = 0.509 (zoom)}).

This smaller portion of Fig. (\ref{k = 0.509}) has its own central island with a chain of smaller ones around it, similar to its larger-scale version (Fig. (\ref{k = 0.509 (zoom)})). The second map figure thus represents analogous information in comparison with the first one, showing the fractal behavior of the standard map. Further successive amplifications would confirm the fractal structure \cite{lichtenberg}.

\section{The Non-Twist Map}

The so-called \emph{Standard Non-Twist map} is defined by the following recurrence equations \cite{del-Castillo-Negrete1996}.
\begin{equation}
\label{Non-Twist Map Equations}
\begin{cases}
    y_{n+1} = y_n - b\sin(2\pi x_n)\\
    x_{n+1} = x_n + a(1 - y_{n+1}^2) \ \ \ \ \  \mbox{mod } 1,
\end{cases}
\end{equation}
in which the variables $x$ and $y$, as in the Equations (\ref{Standard Map Equations}), represent the action-angle variables $\theta$ and $\mathcal{J}$ respectively. The two parameters $a,b\in \mathbb{R}$ regulate the map topology.

A non-twist map is defined as a map that violates the twist condition (Equation (\ref{Twist-Condition})) at least in one invariant curve \cite{wurm2005,del-Castillo-Negrete1996,carvalho1992, corso1998}. In this section, we will only work with the standard non-twist map, in which the twist condition is violated (even for $\epsilon = 0$), at $y=0$. 

Distinguishing from the standard map, the standard non-twist map has in the definition Equations (\ref{Non-Twist Map Equations}) a quadratic function $a(1-y_{n+1}^2)$. For $b=0$ the derivative of the quadratic function with respect to $x$ is known as the \emph{shear function} and it provides an extremum point on the winding number profile. The value of $a$, thus, has a direct influence on the nonmonotonicity of the winding number profile for the non-twist map.

In Fig. (\ref{a = 0.35}) we have the standard non-twist map for the parameters $a = b = 0.35$. The curve referent to the invariant that violates the twist condition is called \emph{shearless curve} \cite{wurm2005,del-Castillo-Negrete1996,carvalho2013} and it can be seen enhanced in red in Fig. (\ref{a = 0.35}). In the same figure, this curve is centrally located on the map. As we shall see, its position can be comprehended from its winding number profile. 

Figure (\ref{a = 0.35}) also shows hyperbolic points that connect the periodic orbits from the \emph{twin chains of islands} in the center region. The twin chains are the islands connected by separatrices in Fig. (\ref{a = 0.35}) on both sides of the shearless curve \cite{wurm2005}. 

Figure (\ref{RN a = 0.35}) has the winding number profile of Fig. (\ref{a = 0.35}), being the reference curve the dashed blue line $x = 0$. Each point in a winding number profile shows the winding number of a specific orbit, the red one referent to the shearless curve. From the same figure, it becomes clear, in this case, that the shearless curve point is related to the maximum of the profile.

The fact that the shearless curve refers to the maximum point in Fig. (\ref{RN a = 0.35}) is a direct consequence of the twist condition (Equation (\ref{Twist-Condition})) violation, since the winding number's derivative is
\begin{equation*}
    \pdv{\omega}{y_n} = 0
\end{equation*}
 there. From calculus, we know that if the slope of a continuous function is zero, there is an extremum or inflection on the curve of the function at that point. Since an extremum can be a local maximum or a local minimum, both possibilities represent a shearless curve point in the winding number profile.

In Fig. (\ref{RN a = 0.35}) we can also notice that the winding number profile is clearly nonmonotonic, as we expected. Non-twist maps have a couple of remarkable properties, different from the standard map ones. In the following subsections, we shall explore some of them.

\subsection{Even Reconnection}

When we have a winding number profile with an extremum, the shearless curve point, there will be two orbits with the same winding number, since the curve is nonmonotonic. This fact is a unique property from the non-twist maps since it originates from the violation of the twist condition (Equation (\ref{Twist-Condition})) and it is known as \emph{degeneracy}. 

If the winding number profile starts to decrease its extremum vertically for a change in parameters $a$ and $b$, becoming flat, the periodic orbits at some point will have the same rotation number (a plateau) \cite{wurm2005}. This sequence of events is called \emph{periodic orbit's collision} and it is followed by its annihilation. 

The periodic orbit's collision involves another process, the reconnection of the separatrices. This reconnection is a process of global bifurcation (large stability changes in an equilibrium \cite{yorke}) that changes the topology of the separatrices, during this event the separatrices collide \cite{wurm2005}. 

The way the reconnection happens depends on the parity of the orbit's period that can be obtained from the number of islands on the map. Notice that the maps are periodic in the $x$-axis, so half islands on it are the same ones from the opposite side. In this section, we take a look at the even-period scenario. 

\begin{figure*}[t]
    \centering
    \begin{subfigure}[b]{.32\textwidth}
        \centering
        \includegraphics[width=\textwidth]{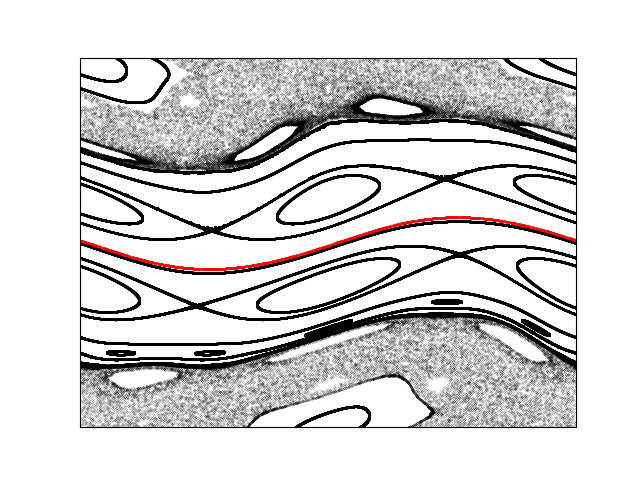}
        \caption{$a = 0.532$ $b = 0.28$.} 
        \label{a = 0.532}
    \end{subfigure}
    \hfill
    \begin{subfigure}[b]{.32\textwidth}
        \centering 
        \includegraphics[width=\textwidth]{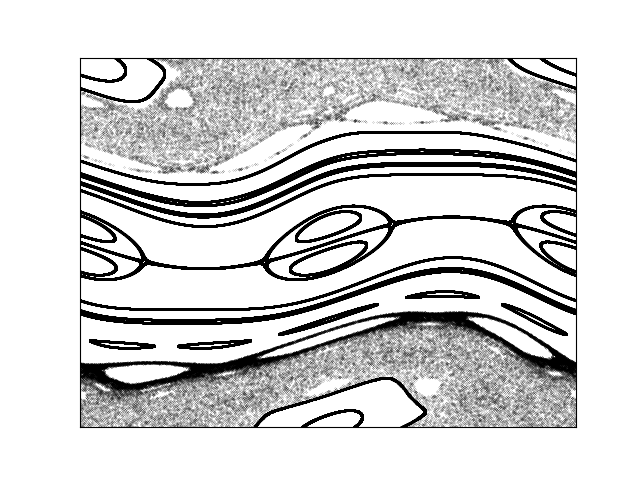}
        \caption{$a = 0.505$ $b = 0.28$.}  
        \label{a = 0.505}
    \end{subfigure}
    \hfill
    \begin{subfigure}[b]{.32\textwidth}
        \centering 
        \includegraphics[width=\textwidth]{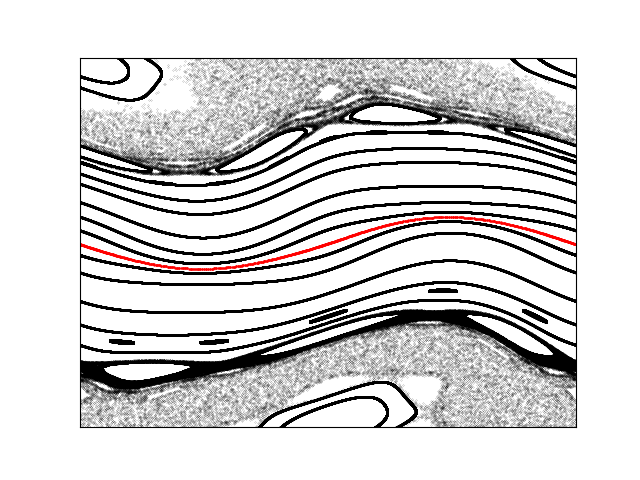}
        \caption{$a = 0.495$ $b = 0.28$.}  
        \label{a = 0.495}
    \end{subfigure}
    \vskip\baselineskip
    \begin{subfigure}[b]{.32\textwidth}
        \centering 
        \includegraphics[width=\textwidth]{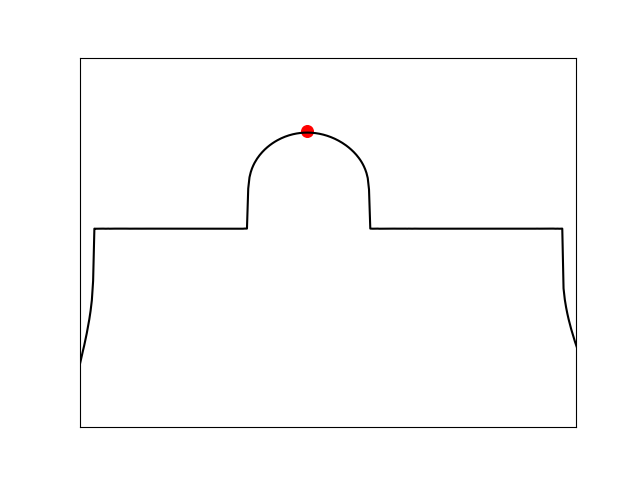}
        \caption{Winding number profile for \\ $a = 0.532$ $b = 0.28$}  
        \label{RN a = 0.532}
    \end{subfigure}
    \hfill
    \begin{subfigure}[b]{.32\textwidth}  
        \centering 
        \includegraphics[width=\textwidth]{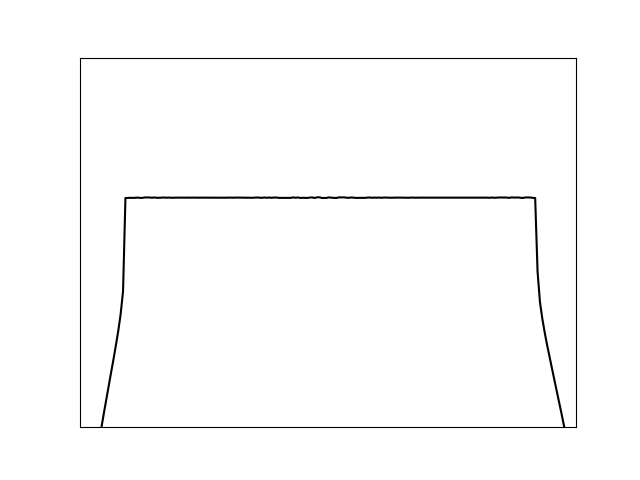}
        \caption{Winding number profile for \\ $a = 0.505$ $b = 0.28$}  
        \label{RN a = 0.505}
    \end{subfigure}
    \hfill
    \begin{subfigure}[b]{.32\textwidth}
        \centering 
        \includegraphics[width=\textwidth]{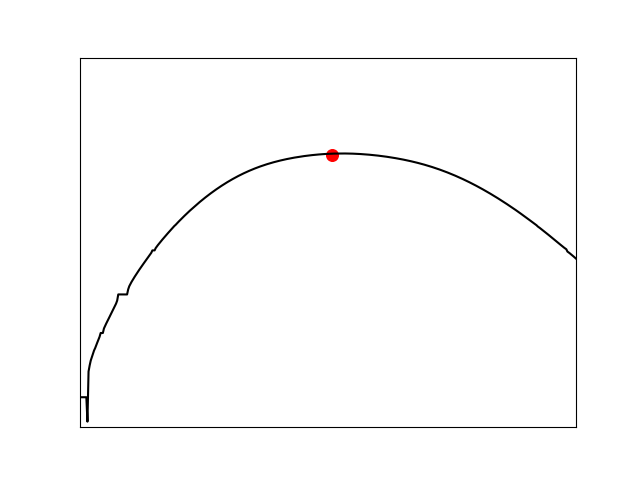}
        \caption{Winding number profile for \\ $a = 0.495$ $b = 0.28$}  
        \label{RN a = 0.495}
    \end{subfigure}
    \caption{Example of even period reconnection sequence of maps (upper row $y$ vs $x$) together with their respective winding number profiles (upper row $\omega$ vs $y$), referenced as WNP.}
    \label{even reconnection}
\end{figure*}

Fig. (\ref{even reconnection}) shows an example of an even reconnection sequence. Initially, in Fig. (\ref{a = 0.532}), we have two twin island chains. Decreasing the parameter $a$ give us Fig. (\ref{a = 0.505}). Now the twin island chains start to merge in a dipole topology \cite{scholarpediaWurm}, the separatrices get closer to each other, becoming one and the islands and hyperbolic points collide. In the same image, the shearless curve vanished during the reconnection process.

Decreasing again the value of $a$ even more, we see that the islands finally collided, vanishing the twin island chains (Fig. (\ref{a = 0.495})). The center has

 \noindent become indistinguishable compared with the first image and the shearless curve has reappeared. 

 From a winding number profile perspective, the  maximum point from the profile in Fig. (\ref{RN a = 0.532}) starts to lower with the decrease of the parameter $a$ as we can see from Fig. (\ref{RN a = 0.532}) to Fig. (\ref{RN a = 0.505}), becoming flat (no extremum, implying the absence of a shearless curve). The last profile Fig. (\ref{RN a = 0.495}), with the further decrease of $a$, shows a new maximum showing the shearless curve on the map once more.
 
 \subsection{Odd Reconnection}

\begin{figure*}[t]
    \centering
    \begin{subfigure}[b]{.32\textwidth}
        \centering
        \includegraphics[width=\textwidth]{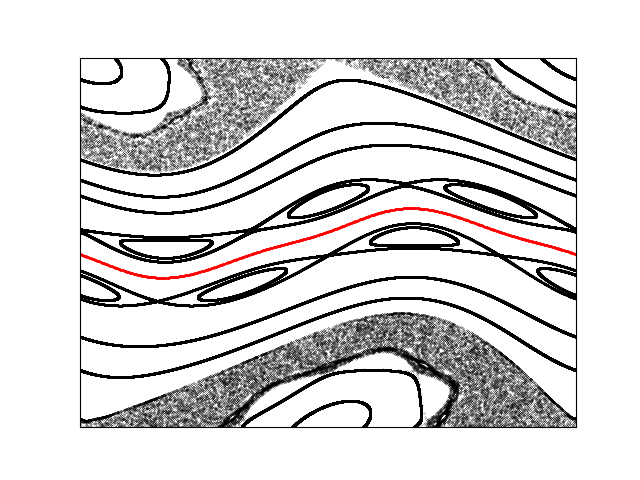}
        \caption{$a = 0.346\ b = 0.30$.} 
        \label{a = 0.346}
    \end{subfigure}
    \hfill
    \begin{subfigure}[b]{.32\textwidth}
        \centering 
        \includegraphics[width=\textwidth]{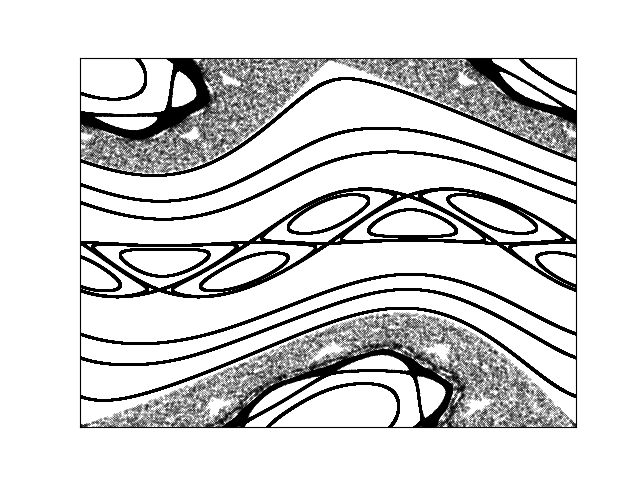}
        \caption{$a = 0.3407\ b = 0.30$.}  
        \label{a = 0.3407}
    \end{subfigure}
    \hfill
    \begin{subfigure}[b]{.32\textwidth}
        \centering 
        \includegraphics[width=\textwidth]{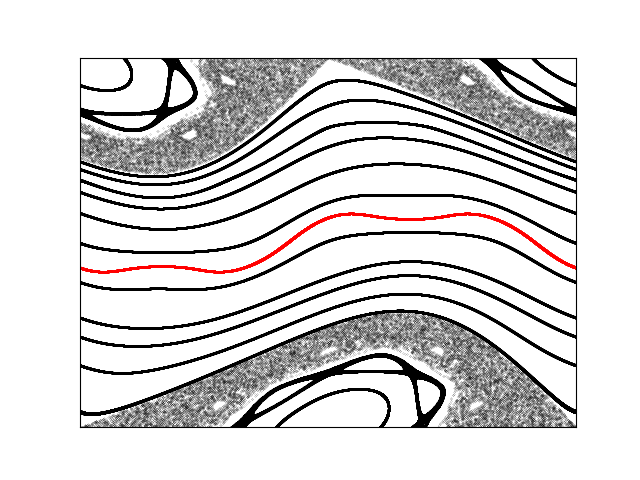}
        \caption{$a = 0.3355\ b = 0.30$.}  
        \label{a = 0.3382}
    \end{subfigure}
    \vskip\baselineskip
    \begin{subfigure}[b]{.32\textwidth}
        \centering 
        \includegraphics[width=\textwidth]{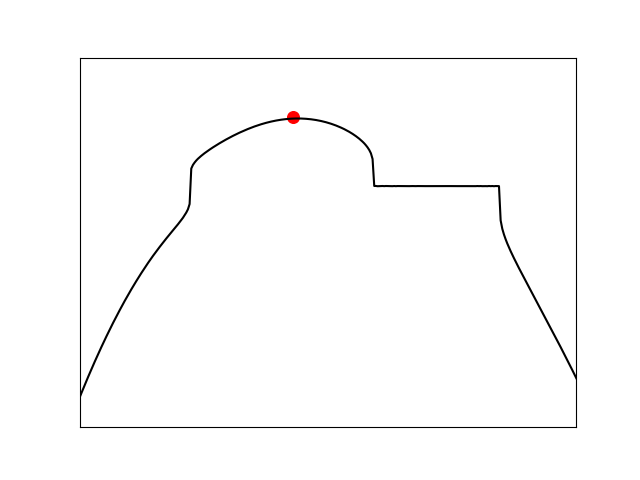}
        \caption{Winding number profile for $a = 0.346$\ $b = 0.30$} 
        \label{RN a = 0.346}
    \end{subfigure}
    \hfill
    \begin{subfigure}[b]{.32\textwidth}  
        \centering 
        \includegraphics[width=\textwidth]{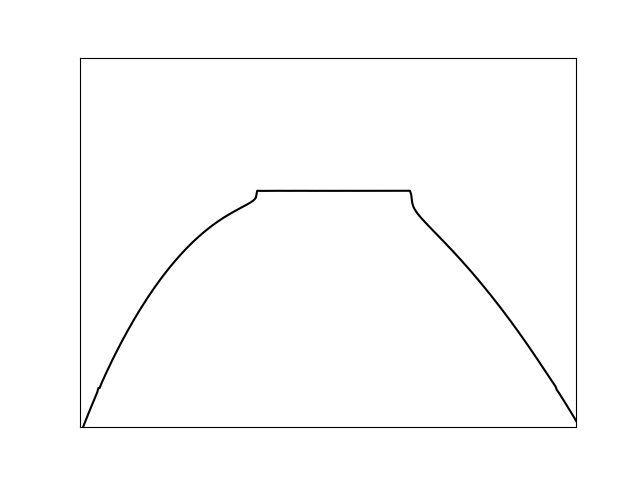}
        \caption{Winding number profile for $a = 0.3407$ $b = 0.30$} 
        \label{RN a = 0.3407}
    \end{subfigure}
    \hfill
    \begin{subfigure}[b]{.32\textwidth}
        \centering 
        \includegraphics[width=\textwidth]{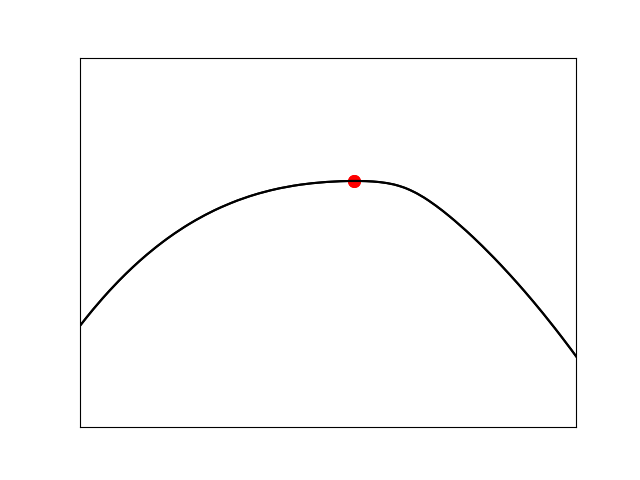}
        \caption{Winding number profile for $a = 0.3355$ $b = 0.30$} 
        \label{RN a = 0.3382}
    \end{subfigure}
    \caption{Example of odd period reconnection sequence of maps (upper row $y$ vs $x$) together with their respective winding number profiles (upper row $\omega$ vs $y$), referenced as WNP.}
    \label{odd reconnection}
\end{figure*}

In a correspondingly way, for an odd period scenario, the reconnection process occurs. However, its development is different from the even period's case as we shall see. As it is evident from Fig. (\ref{a = 0.346}), we have an odd number of islands in each chain, setting an odd reconnection scenario. We decrease the parameter $a$ originating Fig. (\ref{a = 0.3407}). Now the shearless curve has already vanished and the separatrices have connected themselves originating a contour-invariant curve around the islands.

If we further decrease the value of $a$, the separatrices from Fig. (\ref{a = 0.3407}) take a meandering shape around the islands until they are suppressed as in Fig. (\ref{a = 0.3382}). The separatrices, forming a \emph{homoclinic} behavior (joins saddle equilibrium points) in the middle, are then replaced by the \emph{heteroclinic} (connects different equilibrium points) curve exhibited in Fig. (\ref{a = 0.3382}) in which the shearless curve shows up again.

From a winding number profile perspective, similarly to the even reconnection, from Fig. (\ref{RN a = 0.346}) to Fig. (\ref{RN a = 0.3407}) the maximum point becomes a plateau, and in the subsequent step (Fig. (\ref{RN a = 0.3382})) the reinstate of the shearless curve comes from the arising of a new maximum.

\begin{figure*}[t]
\centering
\begin{subfigure}[b]{.49\textwidth}
    \centering
    \includegraphics[width=\textwidth]{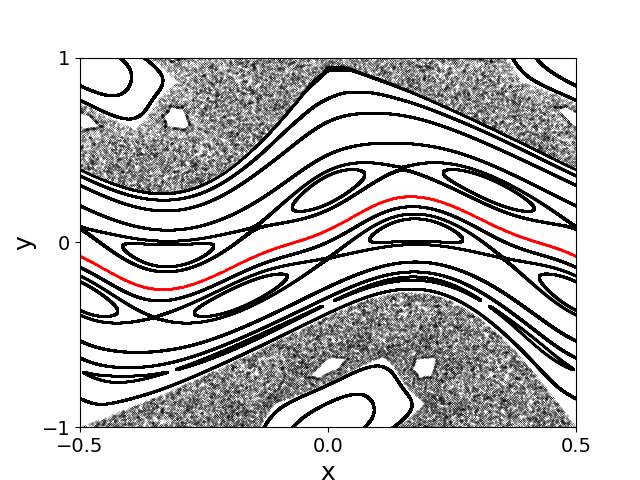}
    \caption{$a = 0.354$ $b = 0.4$.} 
    \label{b = 0.4}
\end{subfigure}
\hfill
\begin{subfigure}[b]{.49\textwidth}
    \centering 
    \includegraphics[width=\textwidth]{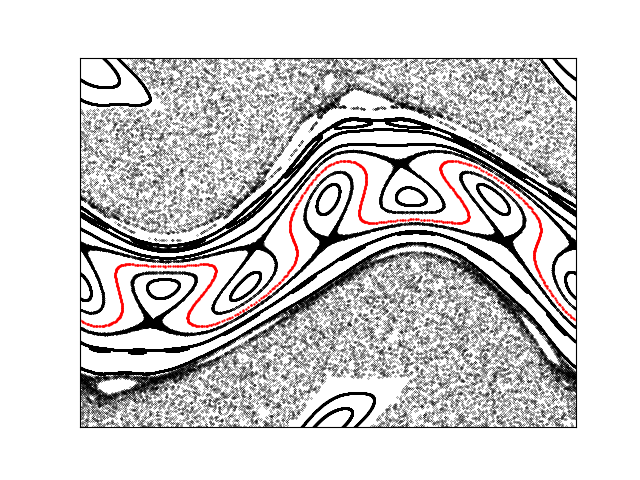}
    \caption{$a = 0.354$ $b = 0.56$.}  
    \label{b = 0.56}
\end{subfigure}
\vskip\baselineskip
\begin{subfigure}[b]{.49\textwidth}
    \centering 
    \includegraphics[width=\textwidth]{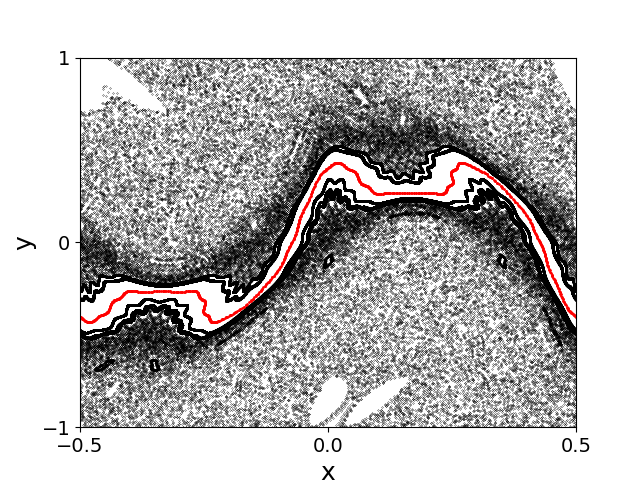}
    \caption{$a = 0.354$ $b = 0.7$.}  
    \label{b = 0.7}
\end{subfigure}
\hfill
\begin{subfigure}[b]{.49\textwidth}  
    \centering 
    \includegraphics[width=\textwidth]{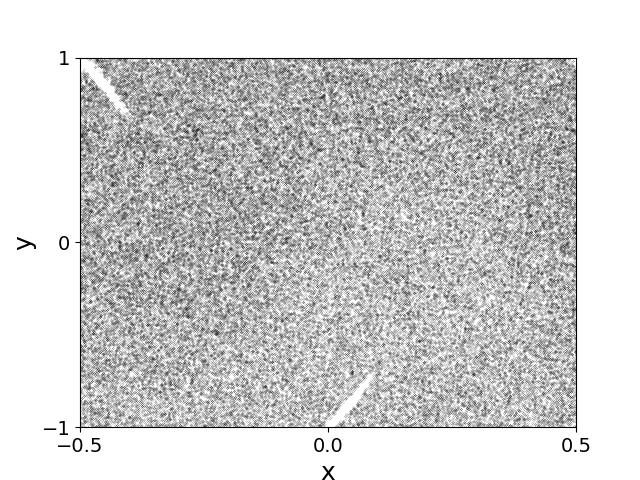}
    \caption{$a = 0.354$ $b = 0.92$}   
    \label{b = 0.92}
\end{subfigure}
\caption{Sequence of maps serving as an example of the shearless curve destruction process.}
\label{shearless curve destruction}
\end{figure*}

\subsection{Shearless Curve Destruction}

Looking at the standard non-twist map definition, Equations (\ref{Non-Twist Map Equations}), the perturbation term $b$ evidently has an important role in the disturbance of the map. If we input a large value for $b$, the number of deformed and destroyed tori will increase expressing regular and chaotic regions \cite{del-Castillo-Negrete1996,del-Castillo-Negrete1997}. Eventually, with the increase of the perturbation in the map, the shearless curve can be destroyed. 

In this section, we shall dissect this Shearless curve breakup process step by step. Starting from similar parameters in comparison with Fig. (\ref{a = 0.346}), we have Fig. (\ref{b = 0.4}). Now, instead of changing $a$ as we did to show the odd reconnection scenario, in order to grow disturbance in the map, we increase $b$. Within an odd reconnection process, Fig. (\ref{b = 0.56}) shows the standard non-twist map for a larger parameter $b$. On the image, we can see that the chaotic regions of the map have grown and the KAM curves and periodic orbits within the separatrices start to reconnect. 

Increasing the perturbation parameter to $b = 0.7$ brings to Fig. (\ref{b = 0.7}). From this point, there are no longer invariant curves, islands, and separatrices, instead, we have a darker region of \emph{stickiness}, around the still-standing shearless curve. Stickiness is the temporary confinement of chaotic orbits in a specific region of the phase space before they diffuse to a larger region after, usually, a large number of iterations \cite{contopoulos2010,szezech2009}. 

Even with the vanishment of the majority of the KAM tori in the center region of the map (\ref{b = 0.7}), the shearless curve is kept intact. This illustrates the shearless curve's resilience to perturbation, being it the most resistant torus \cite{abud2013,del-Castillo-Negrete1996,scholarpediaWurm}. In the rupture limit to global chaos, the shearless torus acts as a \emph{transport barrier}. This name comes from one of its traits: the barrier prevents transport between the chaotic regions \cite{szezech2009,fonseca2014}. Even after the shearless curve destruction, the stickiness remains, keeping the transport barrier a little longer \cite{szezech2009}.

In the last image of the process (Fig. (\ref{b = 0.92})), the chaotic sea takes over the last torus (the shearless curve) and the darker stickiness region becomes homogeneous chaos. This expresses the shearless curve destruction, also known as \emph{Transition to Global Chaos in Non-Twist Maps} \cite{del-Castillo-Negrete1996}. Thus by increasing the parameter $b$, the perturbation on the map increases. Since this parameter is directed related to the disturbance parameter $\epsilon$ in Equation (\ref{Perturbed Hamiltonian}), for $b\gg 1$, KAM's theorem implies a larger number of chaotic tori, the same behavior exposed in Fig. (\ref{odd reconnection}).

\section{Parameter Space}

The shearless breakup shown in Fig. (\ref{shearless curve destruction}) occurs for a specific set of parameters $(a,b)$. In order to understand which ones lead to this scenario, there is a method proposed by Shinohara and Aizawa \cite{shinohara1997}, using the \emph{indicator points} to create a \emph{parameter space} or \emph{shearless breakup diagram}. Indicator points can be generally described as a set of four points
\begin{equation}
    \label{Indicator Points}
    z_1^{(\pm)} = \left(\pm \frac{1}{4},\pm \frac{b}{2}\right), \ \ z_2^{(\pm)} = \left(\frac{a}{2}\pm\frac{1}{4},0\right),
\end{equation}
that necessarily are contained in the shearless curve \cite{shinohara1998}, if it exists.

The parameter space of the shearless curve

\begin{figure}[htb]
    \centering
    \includegraphics[width=0.7\textwidth]{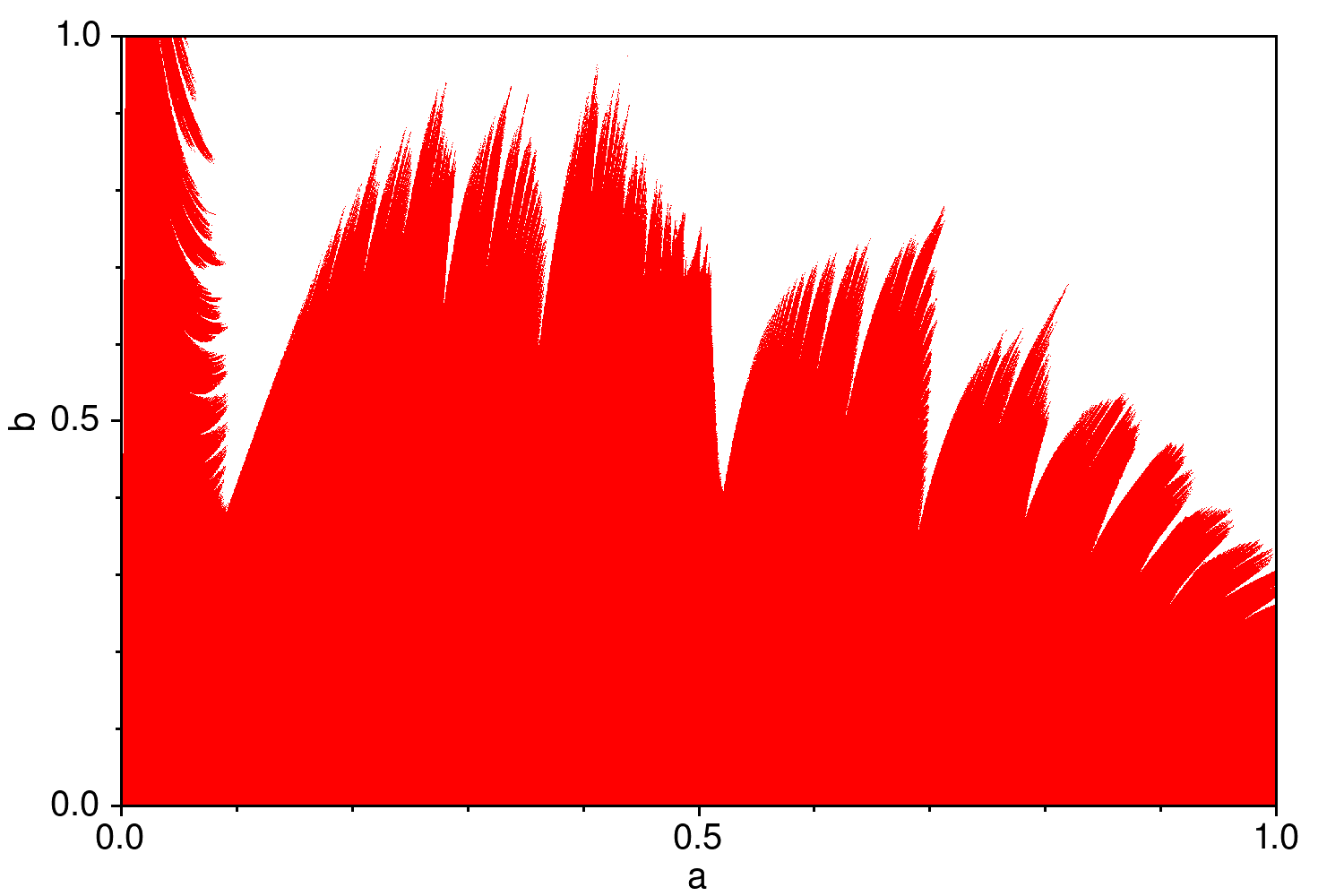}
    \caption{Shearless breakup diagram for a mesh of $2000\times 2000$ points, intervals $0<a<1$ and $0<b<1$, threshold range $\abs{y} < 10$ and $10^6$ iterations.}
    \label{Parameter Space}
\end{figure}

\noindent consists of a mesh of points whose axis are the parameters $a$ (x-axis) and $b$ (y-axis), usually limited to the same interval. One way of building a parameter space is checking, for a large number of iterations if the $y$ value of the indicator points for a determined point $(a,b)$ is restrained to a specified threshold range. 
 
If we set the points inside the threshold range to be colored red and the ones outside white, the resulting graphic is an approximation of the parameter space in the limited interval. For a mesh of $2000\times 2000$ points, the intervals $0<a<1$, $0<b<1$ and the threshold range $\abs{y} < 10$, Fig. (\ref{Parameter Space}) shows the resulting parameter space after $10^6$ iterations of the indicator points. 

The white points (out of the threshold range) represent the parameters $(a,b)$ for which the shearless curve has been broken, while in the red ones it is maintained. Fig. (\ref{Parameter Space}) shows that the points for which the shearless curve does exist and the ones it does not are separated by a sharped boundary. Taking a closer look at Fig. (\ref{Parameter Space}) we can visualize fractal and continuous boundaries \cite{mathias2019,mugnaine2020}.

From Fig. (\ref{Parameter Space}), when $b$ increases, it is noticeable the shearless breakup tendency grows. However, this can be compensated by decreasing $a$. Thus, the parameter space acts as a useful tool in order to investigate which set of parameters $(a,b)$ preserves the shearless curve and to visualize how a change in the value of one of them or both influences the possibility of this scenario. The continuous boundaries of the parameter space are related to the collision of the twin islands \cite{wurm2005}.

\section{Conclusion}

In this review, we introduced the standard map and the non-twist standard map. These maps can be derived as approximations of Poincaré maps of Hamiltonian of quasi integrable systems of two degrees of freedom. Invariant curves, islands, and chaotic trajectories are introduced. The route to global chaos in phase space is shown by varying control parameters. Special features of the standard non-twist systems contained in the non-twist map are presented. Among these features, are the onset and break up of shearless invariants curves, as well as their dependence on the control parameters.

\section*{Acknowledgments}

We acknowledge the financial support from the scientific agencies: São Paulo Research Foundation (FAPESP) under Grant Nos. 2022/04251-7, 2022/05667-2 and 2018/03211–6, and Conselho Nacional de Desenvolvimento Cientíıfico e Tecnológico (CNPq) under Grant No. 304616/2021-4.

\printbibliography

\end{document}